\documentclass{elsart}
\usepackage{amsmath}
\newcommand{\bm}[1]{\mbox{\boldmath$#1$}}
\journal{Physics Letters}
\begin{document}
\begin{frontmatter}

\title{Inverse problem for the retarded field of an arbitrary moving
charge}
\author{V.Ya. Epp\corauthref{cor1}},
\corauth[cor1]{Corresponding author.}
\ead{epp@tspu.edu.ru}
\author{T.G. Mitrofanova}
\address{Tomsk State Pedagogical University, 634041 Tomsk, Russia}

\begin{abstract}
It is assumed that the Lienard-Wiechert fields of an
arbitrary moving charge is measured or predefined as a function of
time. The position of the charge is calculated as a function of the
retarded time.
\end{abstract}

\end{frontmatter}

\section{Introduction.}
The electric and magnetic fields of a point-like charge $e$ moving
along the trajectory described by radius vector $\bm r'(t')$ is given
by the Lienard-Wiechert formula \cite{1,14}
\begin{eqnarray}\label{13}
\bm E(t)&=&\frac{e(\bm n-\bm\beta)(1-\beta^2)}{R^2(1-\bm\beta\cdot\bm
n)^3}+\frac{e}{Rc^2}\frac{\bm n\times[(\bm n-\bm\beta)
\times\bm a]}{(1-\bm\beta\cdot\bm n)^3},\\
\label{13a}
\bm H(t)&=&\bm n\times\bm E(t),
\end{eqnarray}
where $\bm n=\bm R/R$, $\bm R=\bm R(t')=\bm r-\bm r'(t')$, $t'$ is the
retarded time
\begin{eqnarray}\label{a3} t=t'+\frac{|{\bm R}(t')|}{c},
\end{eqnarray}
$\bm\beta=d\bm r'/c dt'$ is the charge velocity in units of speed
of light $c$, $\bm a=d^2\bm r'/{dt'}^2$ is  the charge acceleration and
$\bm r$ is the observer position.

We state the inverse problem as follows. The fields $\bm E(t)$ and $\bm
H(t)$ are the known functions of time. It is required to calculate the
position $\bm r'(t')$ of the charge.

A particular case of this problem is the inverse problem of radiation.
In this case only the last term of expression (\ref{13}) is taken into
account. The inverse problem of radiation is of great practical
significance. There are two most actual aspects of the problem: i)
calculation of the charge trajectory from its radiation, ii)
construction of a radiation source with preassigned spectral and
polarization properties. This problem has become the topic of a large
body of research. There were for instance suggested some ways to
construct sources with given spectral properties of radiation, based on
an undulator \cite{2,3,5,6}. An alternative approach
concerning radiation in a "short magnet" was demonstrated by authors of
Refs. \cite{7,8}. The general solution of the problem is
presented in Refs. \cite{11,12} (see also \cite{13}).

In present paper we suggest the solution of the inverse problem
taking into account both terms of Eq. (\ref{13}) -- the generalized
Coulomb field and the field of radiation. The fields ${\bm E}(t)$ and
$\bm H(t)$ are the functions of $\bm R(t')$ and its derivatives
$\dot{\bm R}(t')$ and $\ddot{\bm R}(t')$. Thus, it might appear at first
glance that the solution of the inverse problem is the solution of
differential equations. The task becomes more complicated because of
nonlinearity of equations and implicit dependence of the fields on the
retarded time $t'$.

On the other hand we have to define only one unknown vector $\bm R$,
knowing two vector functions ${\bm E}(t)$ and ${\bm H}(t)$. This makes
possible to escape solving the differential equations and to do only
with algebraic transformations. Alone the case $\bm H(t)=0$ and $\bm
E(t)=\bm kE(t)$, $(\bm k=\mbox{const})$ requires integration.

\section{General solution.}
Instead of Eq. (\ref{13}) we use the Feynman's formula for $\bm E(t)$
\cite{9} (another form of it is given in \cite{15}).
\begin{eqnarray}\label{14} \bm E(t)=e\left[\frac{\bm
n}{R^2}+\frac{R}{c}\frac{\partial}{\partial t} \left(\frac{\bm
n}{R^2}\right)+\frac{1}{c^2}\frac{\partial^2\bm n} {\partial
t^2}\right], \end{eqnarray}
It follows from Eq. (\ref{13a}) that the vector $\bm n$ lies in the
plane orthogonal to $\bm H(t)$. The known values of $\bm E$ and $\bm H$
allow us to find the angle between the vectors $\bm n$ and $\bm E$.
Indeed, Eq. (\ref{13a}) gives
\begin{eqnarray}\label{*}
\bm E\cdot\bm n=\pm\sqrt{E^2-H^2}, \end{eqnarray}
The sign in the last equation depends on the sign
of the charge. This is seen from comparison of Eq. (\ref{*}) and a
scalar product derived from Eq. (\ref{13}):  \begin{eqnarray}\label{**}
(\bm n\bm E)=\frac{e}{R^2}\frac
{1-\beta^2}{(1-\bm\beta\cdot\bm n)^2}.\end{eqnarray}
Thus, the sign in Eq. (\ref{*}) is the same as the sign
of the charge.
Taking the vector product of Eq. (\ref{13a}) by $\bm E$
we get
 \begin{eqnarray}\label{17}
\bm n=\frac{\bm E\times\bm H+\dfrac{e}{|e|}\bm
E\sqrt{E^2-H^2}}{E^2}.  \end{eqnarray}
The last equation gives $\bm n$ as a function of time $t$ with an
implicit dependence on the retarded time $t'$ according to Eq.
(\ref{a3}). But up to now we do not know the relation between $t$ and
$t'$.

Let us take the derivative $\partial/\partial t$ in Eq. (\ref{14}).
This gives
\begin{eqnarray}\label{+}
\bm E=e\left[\frac{\bm n}{R^2}+\frac{R}{c}\bm n\frac{\partial}
{\partial t}\frac{1}{R^2}+\frac{1}{Rc}\frac{\partial\bm n}{\partial t}+
\frac{1}{c^2}\frac{\partial^2\bm n} {\partial^2 t}\right].
\end{eqnarray}
Substituting this in Eq. (\ref{13a}) we obtain
\begin{eqnarray}\label{18}
\bm H=\frac{e}{cR}\bm n\times\dot{\bm n}+
\frac{e}{c^2}\bm n\times\ddot{\bm n}.
\end{eqnarray}
Points denote the derivatives with respect to time $t$, $\bm n$ is a
known function given by the Eq. (\ref{17}). Multiplying the last
equation by $\bm H$ we find
\begin{eqnarray}\label{19}
H^2=\frac{e}{cR}\bm H\cdot(\bm n\times\dot{\bm n})+
\frac{e}{c^2}\bm H\cdot(\bm n\times\ddot{\bm n}).
\end{eqnarray}
This gives the length of the radius vector $R$
\begin{eqnarray}\label{21}
R=\frac{ec\bm H\cdot(\bm n\times\dot{\bm n})}
{c^2H^2-e\bm H\cdot(\bm n\times\ddot{\bm n})}
\end{eqnarray}
at the retarded time $t'$. But the right-hand side of Eq. (\ref{21}) is
given as the function of $t$. Substituting $R$ into Eq. (\ref{a3}) we
find the retarded time $t'$
\begin{eqnarray}t'=t-\frac{R}{c}.\notag
\end{eqnarray}
Multiplying Eq. (\ref{21}) by vector $\bm n$ we obtain finally
\begin{eqnarray}\label{23}
\bm R=\bm n\frac{ec(\bm H\cdot\bm N)}
{c^2H^2-e(\bm H\cdot\dot{\bm N})},
\end{eqnarray}
where $\bm N=\bm n\times\dot{\bm n}$ with $\bm n$ given by Eq.
(\ref{17}).

This is the general solution of the problem in case of $\bm H(t)\neq 0$.

\section{Singularities.}

Let us consider the case $\bm H(t)=0$. We can rewrite Eq.
(\ref{18}) as follows
\begin{eqnarray}\label{L}
\bm n\times(\dot{\bm n}+\frac{R}{c}\ddot{\bm n})=0.
\end{eqnarray}
We assume first that $\ddot{\bm n}\neq 0$, hence $\dot{\bm n}\neq 0$.
Then Eq. (\ref{L})  means that all of the three vectors $\bm n$,
$\dot{\bm n}$ and $\ddot{\bm n}$ lie in the some plane. Vectors $\bm n$
and $\dot{\bm n}$ are perpendicular to each other because of identity
$\bm n\cdot\bm n=1$, which gives $\dot{\bm n}\cdot\bm n=0$.
Taking the vector product of $\dot{\bm n}$ and Eq. (\ref{L}) we obtain
\begin{eqnarray}\label{LL}
R(\dot{\bm n}\cdot\ddot{\bm n})+c\dot{\bm n}^2=0.
\end{eqnarray}
Hence
\begin{eqnarray}R=-\frac{c\dot{n}^2}{(\dot{\bm n}\cdot\ddot{\bm n})}.
\notag\end{eqnarray}
Multiplying by $\bm n$ from Eq. (\ref{17}) we find \footnote{It is
seen from Eq. (\ref{**}) that $\bm E\neq 0$.} \begin{eqnarray}\label{M}
\bm R=-\frac{e}{|e|}\frac{c\dot{n}^2}{\dot{\bm n}\cdot\ddot{\bm n}}
\frac{\bm E}{E}.
 \end{eqnarray}

And finally we consider the case $\dot{\bm n}=0$, $\ddot{\bm n}=0$,
i.e. $\bm n=(e\bm E)/(|e|E)=\mbox{const}$. This means that the point
charge is moving along a straight line through the point where the
field is measured. Direction of the line is given by vector $\bm E$.
Position of the charge on the line is defined by the scalar function
$R(t)$ which can be derived from Eq. (\ref{+}). Multiplying it by $\bm
n$ we find
\begin{eqnarray}E(t)=|e|
\left(\frac{1}{R^2}+\frac{R}{c}\frac{d}{dt}\frac{1}{R^2}\right).
\notag\end{eqnarray}
Change of the variable $R=1/x$ gives the well known Riccati equation
\cite{10}
\begin{eqnarray}\label{N}
\dot{x}+\frac{c}{2}x^2=\frac{c}{2|e|}E(t).
 \end{eqnarray}
As an example we calculate position of a charge which produces a
constant electric field $\bm E_0$ at some given point. Eq. (\ref{N})
has three solutions for $E(t)=E_0$:
\begin{eqnarray}\label{str}
\bm R_1&=&\frac{e\bm E_0}{|eE_0|^{3/2}}\notag\\
\bm R_2&=&\bm R_1\coth\xi t,\quad
t'=t-\frac{R_1}{c}\coth\xi t,\\
\bm R_3&=&\bm R_1 \tanh\xi t,\quad
t'=t-\frac{R_1}{c}\tanh\xi t,\notag
\end{eqnarray}
where $\xi=\frac{c}{2}\sqrt{E_0/|e|}$. The first solution is
consistent with what is expected -- it follows from Coulomb's law for a
point-like charge at rest. The others are in some sense
unexpected. The second one shows that a constant field can be produced
by a charge moving from infinity to an observer with a velocity close
to the speed of light in infinity and slowing down to zero while the
charge approaches the point a distance $R_1$ from the observer. The
third solution represents a charged particle moving from the observer
to the point with coordinates $\bm R_1$ slowing down from speed of light
to zero. Note that the obtained solutions remain valid for a finite
period of time. As long as the charged particle moves according to law
(\ref{str}) it produces a constant electric field at the given point.

Thus, the solution of the inverse problem is given by Eqs. (\ref{17})
and (\ref{23}) in case $\bm H\neq 0$ and Eqs. (\ref{17}) and
(\ref{M}) when $\bm H=0$, $\ddot{\bm n}\neq 0$, $\dot{\bm n}\neq 0$. If
$\bm H=0$, $\ddot{\bm n}=0$, $\dot{\bm n}=0$ then the solution is
represented by Eq. (\ref{17}) and solution of Eq. (\ref{N}). We see
that the same field can be produced by a negative or positive charge.
But the trajectory of the positive charge differs from that
of the negative one.


\begin{thebibliography}{99}
\bibitem[1]{1} L.D. Landau and E.M. Lifshitz, The Classical Theory of
Fields, Pergamon, New York, 1975.
\bibitem[2]{14} D.J. Griffiths, Introduction to Electrodynamics,
 Prentice Hall, N. Jersey, 1999.
\bibitem[3]{2} M.M. Nikitin and V.Ya. Epp, Sov. Phys. Tech. Phys.
21 (1976) 1404.
\bibitem[4]{3} M.M. Nikitin and N.I. Fedosov, Sov. Phys. Tech. Phys.
22 (1977) 1438.
\bibitem[5]{5} E.G. Bessonov and A.V. Serov, Preprint FIAN No 62,
Lebedev Physics Institute, Moscow, 1982.
\bibitem[6]{6} N.V. Smoljakov and A.I. Chechin, Nucl. Instr. and Meth.
A359 (1995) 97.
\bibitem[7]{7} V.G. Bagrov, M.M. Nikitin, I.M. Ternov, N.I. Fedosov,
Nucl. Instr.  Meth. 208 (1983) 167.
\bibitem[8]{8} V.G. Bagrov,
I.M. Ternov, N.I. Fedosov, Phys. Rev. D28 (1983) 2464.
\bibitem[9]{11} V.G. Bagrov, M.M. Nikitin, V.F. Zal'mezh, N.I. Fedosov
and V.Ya. Epp, Nucl. Instr. and Meth. A239 (1983) 579.
\bibitem[10]{12} V.G.Bagrov, V.F. Zal'mezh, M.M. Nikitin and V.Ya. Epp,
 Nucl. Instr. and Meth. A261 (1987) 54.
\bibitem[11]{13} Synchrotron
Radiation Theory and Its Development.  Ed.  by V.A. Bordovitsyn, World
 Scientific, 1999.
\bibitem[12]{9} R.P. Feynman, R.B. Leighton and M. Sands, The Feynman
Lectures on Physics, Vol. 1, Ch. 28, Eq. (28.3), Addison-Wesley,
Reading MA, 1989.
\bibitem[13]{15} M.A. Heald and J.B. Marion,
Classical Electromagnetic Radiation. Sounders College Publishing,
Orlando, US, 1995.
\bibitem[14]{10} E. Kamke,  Differential
Gleichungen, Leipzig, 1959.
\end{thebibliography}
\end{document}